\documentclass[preprintnumbers, prd, floatfix,twocolumn,preprintnumbers, letterpaper, superscriptaddress,nofootinbib]{revtex4-1}
\usepackage{amsfonts}
\usepackage{amsmath}
\usepackage{amssymb,epsf}
\usepackage{latexsym}
\usepackage{graphicx,epsfig}
\usepackage{amssymb}
\usepackage{float}
\usepackage{subfigure}
\usepackage{epstopdf}
\usepackage[colorlinks=true,citecolor=blue,linkcolor=blue,urlcolor=black]{hyperref}
\usepackage{dcolumn}
\usepackage{psfrag}
\usepackage{wrapfig}
\usepackage{makeidx}
\usepackage{epsf}
\usepackage{color}
\usepackage{multirow}
\usepackage{tabularx}
\usepackage{array}
\usepackage{booktabs}
\newcolumntype{c}[1]{>{\centering\arraybackslash}p{#1}}

\graphicspath{{Images/}}

\usepackage{tikz}
\definecolor{lime}{HTML}{A6CE39}

\begin{document}
\title{Thermodynamical properties of nonsingular universe}
\author{Ahmad Sheykhi}
\email{asheykhi@shirazu.ac.ir}
\address{Department of Physics, College of
Science, Shiraz University, Shiraz 71454, Iran}
\address{Biruni Observatory, College of Science, Shiraz University,
Shiraz 71454, Iran}
\author{Leila Liravi}
\address{Department of Physics, College of
Science, Shiraz University, Shiraz 71454, Iran}
\author{Kimet Jusufi}
\email{kimet.jusufi@unite.edu.mk}
\address{Physics Department,
State University of Tetovo, Ilinden Street nn, 1200, Tetovo, North
Macedonia}

\begin{abstract}
We disclose the thermodynamical properties of the apparent horizon
in a nonsingular universe. We take into account the zero-point
length correction to the gravitational potential and derive the
modified entropy expression that includes zero-point length
correction terms. We apply the first law of thermodynamics on the
apparent horizon as well as the emergent gravity scenario to
derive the modified Friedmann equations. Further, we examine the
time evolution of the total entropy, including the entropy of the
apparent horizon and the matter field entropy inside the horizon
and find out that the generalized second law of thermodynamics is
satisfied. We also investigate the cosmological implications of
the modified cosmology through zero-point length. We observe that
the zero-point length correction does not change the general
profile of the universe evolution, however, it shifts the time of
the phase transition in a universe filled with matter and
cosmological constant. We explore the age of the universe for our
model and observe that the predicted age of the universe becomes
larger compared to the standard cosmology. By calculating the
explicit form of Ricci and Kretchmann invariants, we confirm that
in our model, the initial singularity of the universe is removed.
This is an expected result, because the main motivation for
considering zero-point length correction in the gravitational
potential is to remove singularity at the origin.
\end{abstract}
\maketitle
\section{Introduction\label{Intro}}
The discovery of black holes thermodynamics in 1970's, prompted
physicists to speculate a profound connection between gravity and
thermodynamics. Jacobson was the first who argued that the
Einstein field equation of general relativity is just an equation
of state for the spacetime \cite{Jac,Pad1,Pad2}. The relationship
between gravitational field equations and the first law of
thermodynamics was generalized to other theories, beyond Einstein
gravity, such as $f(R)$ gravity \cite{Eling}, Gauss-Bonnet gravity,
scalar-tensor gravity, and more general Lovelock gravity
\cite{Pad3,Akbar1}. In the cosmological setups, it was shown that
the first law of thermodynamics on the apparent horizon,
$dE=TdS+WdV$, can be translated as the first Friedmann equation of
the Friedmann-Robertson-Walker (FRW) universe
\cite{CaiKim,Wang1,Akbar2,Sh1,Sh2}. Here, $T$ and $S$ are,
respectively, the temperature and entropy associated with the
apparent horizon, $dE$ is the change in the total energy inside
the apparent horizon due to the volume change $dV$ of the
universe, while $W$ is the work density. The connection between
thermodynamics and gravity reveals that the field equations of
gravity are just equation of state for the spacetime.

Another approach for understanding the nature of gravity was
proposed by Verlinde \cite{Ver}, who argued that gravity is not a
fundamental force and can be regarded as an entropic force which
appears due to the change in the information associated with the
system.  Two main assumptions of Verlinde's proposal are the
holographic principle and the equipartition law of energy for the
degrees of freedom of  the system. Using these two principles,
Verlinde derived  Newton's law of gravity, the Poisson equation
and in the relativistic regime, the field equations of general
relativity. The studies on the entropic force scenario were
generalized to cosmology, where it has been shown that the
Friedmann equations can be derived by using Verlinde's proposal
\cite{Cai2,sheyECFE}. It has been shown that Verlinde's proposal
of emergent gravity suffers from some internal inconsistencies
\cite{Dai} (see also \cite{Car}).

Although Jacobson and Verlinde 's arguments reveal the
thermodynamic nature of gravity from  the point of view of
thermodynamics and statistical mechanics, however, they assume
spacetime as a pre-existing geometry. At a deeper level,
Padmanabhan \cite{emer} argued that the spatial expansion of our
Universe can be understood as a consequence of the emergence of
space, and the cosmic space is emergent as the cosmic time
progresses.  By equating the difference between the number of
degrees of freedom in the bulk and on the boundary, Padmanabhan
derived the Friedmann equation that describes the evolution of the
Universe \cite{emer}. The idea that gravity is an emergent
phenomenon not only has been checked in Einstein gravity, but also
has been generalized to Gauss-Bonnet gravity, Lovelock gravity,
and braneworld scenarios \cite{Cai,Yang,Sh3,Sh4,Sh5}. In all these
studies, the aim is to extract the dynamical field equations in
the background of FRW universe, by calculating the number of
degrees of freedom on the boundary and in the bulk.

Despite the thermodynamic nature of gravity, general relativity
suffers some basic problems such as singularity at the center of
the black hole solutions and initial singularity at the origin of
the Universe. The singularity also exists in Newtonian gravity for
a point particle with mass $M$, where the gravitational potential,
$\phi=-GM/r$, diverges at $r=0$. Many attempts have been made to
remove singularities in the solutions of general relativity. For
example, regular charged black holes have been proposed which are
nonsingular local solutions of the field equations in the presence
of nonlinear electrodynamics \cite{RegBH1,RegBH2}. There is also
an independent way for removing singularity in the black hole
theory, which is based on the concept of T-duality in string
theory. According to T-duality, the momentum space propagator for
the massless particles is given
by~\cite{Padm,Sma,Spa,Fon,Nicolini}
\begin{eqnarray}
\label{G2} G(k)=-\frac{l_0 }{\sqrt{k^2}} K_{1}\left(l_0
\sqrt{k^2}\right),
\end{eqnarray}
where $k=p/\hbar$, $l_0$ represents the zero-point length of
spacetime, and $K_{1}(x)$ is a modified Bessel function of the
second kind. Based on this, the static Newtonian potential
corresponding to $G(k)$ at distance $r$ gets modified as
~\cite{Nicolini,Nicolini2,Nicolini3}
\begin{eqnarray}\label{phir}
\label{V} \phi(r)=-\frac{M }{\sqrt{r^2+l_0^2}},
\end{eqnarray}
where $M$ is the mass of the source of the potential and  for
simplicity, we have set $G=\hbar=c=1$. We would like to point out
an interesting similarity between the modified potential in
T-duality and the Plummer potential, a phenomenological model used
in astrophysics that was introduced in 1911
\cite{Plummer:1911zza}. Using the above potential, the metric of
the Schwarzschild black hole is obtained as \cite{Nicolini}
\begin{eqnarray}
g_{tt}(r)=g^{-1}_{rr}(r)=1-\frac{2Mr^2}{\left(r^2+l_0^2\right)^{3/2}},
\end{eqnarray}
which is regular at $r=0$. The main difference between this
solution inspired by T-duality and those obtained for regular
black holes in \cite{RegBH1,RegBH2} is that here we do not need
nonlinear electrodynamics for removing singularity at $r=0$.

It is important to note that models based on the the zero-point
length inherently contain the effect of generalized uncertainty
principle (GUP). In our case, one way to see this fact is to take
the solution in Eq. \cite{Nicolini} and find the location of the
outer horizon (which can be found after we perform a series
expansion around $l_0$) yielding $r_+=2M'=2M-3l_0^2(4M)$. In other
words, this suggests a modification of the mass $M'=M(1+\beta
l_0^2/M^2)$, where $\beta$ is negative in our case $\beta=-3/8$,
while the mass is corrected as $M'=M+\beta l_0^2/M$. Such an extra
correction term proportional to $ M^{-1}$ was found to be also the
case in the context of GUP \cite{Carr:2015nqa}.

In the present work, we would like to examine the cosmological
consequences of zero-point length during the history of the
universe. Recently, the cosmological viability of the modified
Friedmann equations through zero-point length was explored in
\cite{Gaet1}. They also explored several cosmological aspects of
the theory, including the inflationary paradigm under the
slow-roll condition, density perturbations in the linear regime,
power-law spectrum as well as the spectrum of gravitational waves
\cite{Gaet1}. It has been shown that this model is compatible with
observational data \cite{Gaet1}. Our work differs from
\cite{Gaet1} in that we obtain the modified Friedmann equations by
applying the emergence scenario of gravity and check the validity
of the generalized second law of thermodynamics. Besides, we
explore some novel cosmological aspects of this theory which were
not addressed in \cite{Gaet1}. In particular, we shall calculate
the age of the universe for several cases and confirm that this
model is capable of removing the initial cosmic singularity and
reproduce a nonsingular universe.

{While nonsingular cosmological models have been explored in
contexts like loop quantum gravity (LQG), our approach uniquely
ties zero-point length corrections-rooted in string T-duality- to
thermodynamic and emergent gravity frameworks. This yields
distinct observational signatures, such as a slower early-universe
expansion rate (Section V) and a minimum scale factor Eq.
(\ref{min}), without invoking quantum geometry or ad-hoc matter
fields.}

This paper is organized as follows. In the next section, we use
 entropic gravity and obtain the entropy expression with
zero-point length correction terms. Then, by applying the first
law of thermodynamics on the apparent horizon, we derive the
modified Friedmann equations inspired by zero-point length. In
section III, we employ the emergence scenario for the cosmic space
and derive the modified Friedmann equations which coincides
with the result of section II. In section IV, we check the
validity of the generalized second law of thermodynamics. In
section V, we explore cosmological consequences of the model for
various cases. We finish with closing remarks in the last section.
\section{Modified Friedmann Equations through zero-point length \label{Fried}}
As we mentioned in the introduction, the gravitational potential
gets modified due to the zero-point length effects. The modified
gravitational potential $\phi(r)$ of spherical mass $M$ at
distance $r$ is given by Eq. (\ref{phir}). If a point particle
with mass $m$ is placed at distance $r=R$ in this gravitational
potential, it will feel the following force
\begin{equation}\label{F5}
\vec{F}=-m \vec{\nabla}
\phi(r)|_{r=R}=-\frac{Mm}{R^2}\left[1+\frac{l_0^2}{R^2}\right]^{-3/2}
\hat{r}.
\end{equation}
Taking the entropic force proposal into account, the correction
term to the horizon entropy is given by \cite{KS}
\begin{eqnarray}\label{ds}
1+\left(\frac{1}{2 \pi
R}\right)\frac{d{s(A)}}{dR}=\left(1+\frac{l_0^2}{R^2}\right)^{-3/2}.
\end{eqnarray}
The total entropy of the horizon is given by
\begin{eqnarray} \label{S}
 S_{h}=\frac{A}{4}+s(A),
\end{eqnarray}
where the correction term is denoted by $s(A)$. Note that $A=4\pi
R^2$ is the area of the horizon which is the boundary of the
volume $V=4\pi R^3/3$. Combining Eqs. (\ref{ds}) and (\ref{S}), we
find
\begin{eqnarray}\label{dS1}
dS_h=2\pi R \left(1+\frac{l_0^2}{R^2}\right)^{-3/2} dR.
\end{eqnarray}
The explicit form of the total entropy associated with the
horizon, can be obtained by integrating the above equations. We
find
\begin{eqnarray} \label{St}
S_h&=&\pi R^2 \left(1+\frac{l_0^2}{R^2}\right)^{-1/2}+3\pi l_0^2
\left(1+\frac{l_0^2}{R^2}\right)^{-1/2}\nonumber \\  &&-3 \pi
l_0^2 \ln \left(R+\sqrt{R^2+l_0^2}\right).
\end{eqnarray}
In the limiting case where $l_0^2\rightarrow 0$, the area law is
recovered, $S_h=\pi R^2$. We assume our Universe is described by
the line element
\begin{equation}
ds^2={h}_{\mu \nu}dx^{\mu} dx^{\nu}+R^2(t)(d\theta^2+\sin^2\theta
d\phi^2),
\end{equation}
where $R(t)=a(t)r$, $x^0=t, x^1=r$, and $h_{\mu \nu}$=diag $(-1,
a^2/(1-kr^2))$ stands for the two dimensional metric. The spatial
curvature of the universe is described by $k = -1,0, 1$, which
corresponds to open, flat, and closed universes, respectively.
Note that here $(r,\theta,\phi)$ are dimensionless comoving
coordinates. The dynamical apparent horizon, a marginally trapped
surface with vanishing expansion, is determined by the relation
$h^{\mu \nu}\partial_{\mu}R\partial_{\nu}R=0$, which implies that
the vector $\nabla R$ is null on the apparent horizon surface. The
explicit evaluation of the apparent horizon for the FRW universe
gives the apparent horizon radius \cite{Akbar2,SheyLog}
\begin{equation}
\label{radius}
 R=a(t)r=\frac{1}{\sqrt{H^2+k/a^2}},
\end{equation}
where $H=\dot{a}/a$ is the Hubble parameter. The associated
temperature $T=\kappa/2\pi$ with the apparent horizon is defined
through the surface gravity \cite{Akbar2}
\begin{equation}
\label{radius}
 \kappa=\frac{1}{2\sqrt{-h}}\partial _{\mu}(\sqrt{-h}h^{\mu\nu}\partial
 _{\nu}R).
\end{equation}
Thus, the explicit evolution of the temperature at apparent
horizon of FRW universe reads \cite{Akbar2}
\begin{equation}\label{Th}
T_h=\frac{\kappa}{2\pi}=-\frac{1}{2 \pi
R}\left(1-\frac{\dot{R}}{2H R}\right).
\end{equation}
For $\dot{R}<2H R$, the temperature becomes negative. In this case
one should define $T=|\kappa|/2\pi$. We further assume the total
energy momentum tensor is in the form of the perfect fluid,
\begin{equation}\label{T1}
T_{\mu\nu}=(\rho+p)u_{\mu}u_{\nu}+pg_{\mu\nu},
\end{equation}
where $\rho$ and $p$ are, respectively, the energy density and
pressure of the matter field inside the universe. The continuity
equation, which shows the conservation of energy in the background
of FRW universe, is given as usual
\begin{equation}\label{Cont}
\dot{\rho}+3H(\rho+p)=0.
\end{equation}
In an expanding universe, the work is done due to the volume
change. The corresponding work density is given by  \cite{Hay2}
\begin{equation}\label{Work2}
W=-\frac{1}{2} T^{\mu\nu}h_{\mu\nu}=\frac{1}{2}(\rho-p).
\end{equation}
The total energy inside a $3$-sphere is $E=\rho (4\pi R^3/3)$,
where its differential form is given by
\begin{equation}
\label{dE21}
 dE=4\pi R^{2}\rho dR-4\pi H R^{3}(\rho+p) dt,
\end{equation}
where we have used the continuity equation (\ref{Cont}). Combining
Eqs. (\ref{dS1}), (\ref{Th}), (\ref{Work2}), (\ref{dE21}) with the
first law of thermodynamics on the apparent horizon, $dE = T_h
dS_h + WdV$, we arrive at
\begin{equation} \label{Fried1}
4\pi H R^3 (\rho+p)dt=\left(1+\frac{l_0^2}{R^2}\right)^{-3/2} dR.
\end{equation}
The above equation can be rewritten as
\begin{equation} \label{Fried2}
-\frac{2}{R^3}\left(1+\frac{l_0^2}{R^2}\right)^{-3/2} dR=
 \frac{8\pi}{3}d\rho,
\end{equation}
where we have again used Eq. (\ref{Cont}). Integrating and then
expanding, for small value of $l_0^2/R^2$, we reach
\begin{equation} \label{Fried3}
-\frac{2}{l_0^2}\left[ 1-
\frac{l_0^2}{2R^2}+\frac{3}{8}\frac{l_0^4}{R^4}+...\right]+C=\frac{8\pi}{3}\rho,
\end{equation}
where $C$ is a constant of integration. When $l_0^2/R^2
\rightarrow 0$, we expect the above relation to restore the
standard Friedmann equation. This implies that
\begin{equation}
C\equiv \frac{2}{l_0^2}-\frac{\Lambda}{3},
\end{equation}
where $\Lambda$ is a constant which can be interpreted as the
cosmological constant. Note that one may also define $C= 2/l_0^2$
by absorbing/neglecting the contribution from vacuum energy to the
total energy density $\rho$.

The final form of the modified Friedmann equation, due to the zero
point length correction terms, is obtained as
\begin{equation} \label{Fried4}
H^2+\frac{k}{a^2}-\alpha
\left(H^2+\frac{k}{a^2}\right)^2=\frac{8\pi}{3}(\rho+\rho_{\Lambda}),
\end{equation}
where $\alpha=3l_0^2/4$, $\rho_{\Lambda}=\Lambda/8\pi$ and we have
used the relation (\ref{radius}). Since $l_0^2$ is very small, we
have neglected the higher order term of $\alpha$ in Eq
(\ref{Fried4}). {If we define  $X=H^2+{k}/{a^2}$, we can rewrite
Eq. (\ref{Fried4}) as
\begin{equation} \label{Friedman41}
 -\alpha X^2+ X- \frac{8\pi}{3}\rho=0,
\end{equation}
where we have set $\rho_{\Lambda}=0=\Lambda$, for simplicity.
Solving Eq. (\ref{Friedman41}), yields
\begin{equation} \label{Friedman42}
X=\frac{2}{3  l_0^2}\left[1\mp\sqrt{1- 8\pi l_0^2\rho}\right].
\end{equation}
In order to arrive at the standard Friedmann equation in the
limiting case where  $l_0\rightarrow 0$, we must choose the minus
sign in the above equation. Expanding Eq. (\ref{Friedman42}), we
arrive at
\begin{equation} \label{Friedman43}
H^2+\frac{k}{a^2}=\frac{2}{3 l_0^2}\left[4\pi l_0^2\rho+8\pi^2
l_0^4\rho^2+...\right].
\end{equation}
If we define $\Gamma=2\pi l_0^2$, we can write the final form as
\begin{equation} \label{Fried5}
H^2+\frac{k}{a^2}=\frac{8\pi}{3}\rho\,(1+\Gamma\rho).
\end{equation}}
As an additional point of information, we note that the modified
Friedmann equation given by the last equation has a form similar
to those found in Brane-cosmology models (see, in particular, Eqs.
(656) and (657) in \cite{Clifton:2011jh}). In such a case, the
factor $\Gamma$ can be linked to the brane tension $\sigma$, by
the inverse relation $\Gamma \sim 1/(2\sigma)$. Another important
result can be found from Eq. \eqref{Fried4}, namely if we set
$\dot{a}=0$ ($H=0$) and $\Lambda=0$; and we further solve this
equation using $a=a_{\rm min}$, we get two branches
\begin{eqnarray}
\frac{k}{a_{\rm min}^2}=\frac{3 \pm \sqrt{9-96 \alpha \pi \rho}}{6 \alpha},
\end{eqnarray}
which implies a critical density to exist with the condition,
i.e., $9-96 \alpha \pi \rho\geq 0|_{\rho=\rho_{\rm c}}$, which
gives $\rho_{\rm c}\leq 1/(8 \pi l_0^2)$. If we set
$\rho=\rho_{\rm c}$ and we solve for the minimal scalar factor we
get
\begin{eqnarray}\label{min}
 a_{\rm min}=\frac{\sqrt{6\,k}}{2}\,l_0.
\end{eqnarray}
This equation indicates the existence of a minimum length for the
scale factor and suggests a possible bouncing scenario for a
closed universe, given the additional condition $\dot{H}>0$. On
the other hand, for a flat universe ($k=0$), Eq. (\ref{Fried5})
reads
\begin{equation} \label{Fried6}
H^2-\alpha H^4=\frac{8\pi}{3}(\rho+\rho_{\Lambda}).
\end{equation}
In the absence of the cosmological constant ($\Lambda=0$), we have
\begin{equation} \label{Fried7}
H^2-\alpha H^4=\frac{8\pi}{3}\rho,
\end{equation}
where $\rho$ is the total energy density. In terms of the density
parameter, we can rewrite the above equation as
\begin{equation} \label{Frie8}
1-\alpha H^2=\Omega,
\end{equation}
where $\Omega=8\pi \rho/(3H^2)=\Sigma_{i} \Omega_{i}$ is the total
density parameter and $ \Omega_{i}=8\pi \rho_{i}/(3H^2)$. It is
clear that in this case $\Omega<1$.

It is instructive to take a glance at Eq. (\ref{Fried5}) in the
high energy limit (early universe) where $\Gamma\rho^2\gg \rho$.
For flat universe ($k=0$) we have
\begin{equation} \label{Fried55}
H^2\approx \frac{8\pi}{3}\Gamma\rho ^2.
\end{equation}
 Using $p=\rho \omega$ in the continuity equation (\ref{Cont}), the following expression is
 obtained,
 \begin{align}\label{rho}
 \rho(t)= \rho{_0}\,  a^{-3(1+\omega)}.
 \end{align}
In the radiation dominated era where $\omega=1/3$, we have
$\rho_{r}=\rho_{r,0} a^{-4}$. In this case Eq. (\ref{Fried55})
admits a solution of the form $a(t)=C_1 t^{1/4}$, with
$C_1\equiv\left[ 4 \sqrt{\frac{8\pi \Gamma}{3}} \rho_{r,0}
\right]^{1/4}$. Thus in this case the scale factor evolves as
$a(t) \sim t^{1/4}$. Comparing to the radiation dominated era in
standard cosmology ($a(t) \sim t^{1/2}$), we see that the rate of
the universe expansion is slower in this case. In a matter
dominated universe where $\omega=0$, we get
 $\rho_{m}=\rho_{m,0} a^{-3}$. Thus, Eq. (\ref{Fried55})
admits the solution $a(t)=C_2 t^{1/3}$, where in this case
$C_2\equiv\left[3 \sqrt{\frac{8\pi \Gamma}{3}} \rho_{m,0}
\right]^{1/3}$. Therefore, the scale factor evolves as $a(t) \sim
t^{1/3}$, which compared to the matter dominated era in standard
cosmology ($a(t) \sim t^{2/3}$), we observe that the rate of our
Universe expansion is slower when the zero point length is taking
into account in the cosmological field equations.
\section{Emergence of the cosmic space}
The notion that gravity is an emergence phenomenon and the cosmic
expansion can be understood by calculating the difference between
degrees of freedom on the horizon and in the bulk was pointed out
by Padmanabhan \cite{emer}. According to Padmanabhan the increase
$dV$ of the cosmic volume, is given by \cite{emer}
\begin{equation} \label{dV}
\frac{dV}{dt}=G(N_{\mathrm{sur}}-N_{\mathrm{bulk}}),
\end{equation}
where $N_{\mathrm{sur}}$ and $N_{\mathrm{bulk}}$ stand for the
number of degrees of freedom on the boundary and in the bulk,
respectively. The studies on the emergence gravity were
generalized to Gauss-Bonnet and Lovelock gravity
\cite{Cai,Yang,Sh3}, and braneworld scenarios \cite{Sh4,Sh5},
where it has been shown that the corresponding Friedmann equations
can be derived by properly modifying proposal (\ref{dV}). In
general, for a nonflat universe, one should modify proposal
(\ref{dV}). The modified version of Padmanabhan's proposal is
given by \cite{Sh3}
\begin{equation}
\frac{dV}{dt}=RH \left(N_{\mathrm{sur}}-N_{\mathrm{bulk}}\right).
\label{dV1}
\end{equation}
where we have set $G=1$ for simplicity. In a flat universe,
$R=H^{-1}$ and Eq. (\ref{dV1}) restores (\ref{dV}). We propose the
temperature associated with the apparent horizon is given by
\begin{equation}\label{T2}
T=\frac{1}{2\pi R}.
 \end{equation}
In order to write down the number of degrees of freedom on the
horizon, we first expand the modified entropy (\ref{St}) in terms
of $l_0 ^2/R^2$. We find
\begin{eqnarray} \label{Sexp}
S_h&=&\pi R^2 +\frac{5l_0^2}{2} -\frac{9\pi}{4} \frac{
l_0^4}{R^2}+....
\end{eqnarray}
Neglecting the term of order $l_0 ^4$, and taking into account the
fact that $N_{\mathrm{sur}}\sim4 S_h$, we can recognize
\begin{eqnarray} \label{Nsur2}
N_{\mathrm{sur}}=4\pi R^2+10 \pi l_0 ^2,
\end{eqnarray}
where the second term comes from the leading order term in the
entropy expansion and reflects the influences of the zero-point
length corrections in the entropy expression.

The Komar energy inside the sphere with volume $ V=4 \pi R^3/3$ is given by
\begin{equation} \label{Komar}
E_{\mathrm{Komar}}=|(\rho +3p)|V.
\end{equation}
From the equipartition law of energy, the number of degrees of
freedom of the matter field in the bulk is given by
\begin{equation}
N_{\mathrm{bulk}}=\frac{2|E_{\mathrm{Komar}}|}{T}.  \label{Nbulk}
\end{equation}
Therefore, the number of degrees of freedom in the bulk is
obtained as
\begin{equation}
N_{\rm bulk}=-\frac{16 \pi^2}{3}  R^4 (\rho+3p), \label{Nbulk}
\end{equation}
where we have assumed $\rho+3p<0$ in an expanding universe.
Combining relations (\ref{Nsur2}) and (\ref{Nbulk}) with Eq.
(\ref{dV1}), after simplifying, we get
\begin{eqnarray}
2 R^{-3}\frac{\dot{R}}{H}-2R^{-2}-5 l_0 ^2 R^{-4}=\frac{8\pi
}{3}(\rho+3p). \label{Frgb11}
\end{eqnarray}
Multiplying both side of Eq. (\ref{Frgb11}) by factor $\dot{a}a$,
after using the continuity equation (\ref{Cont}), we arrive at
\begin{equation}\label{Frgb2}
\frac{d}{dt}\left( a^2 R^{-2}\right)+\frac{5}{2}\frac{l_0
^2}{R^4}\frac{d}{dt} (a^{2})=\frac{8 \pi }{3} \frac{d}{dt}(\rho
a^2).
\end{equation}
\\
Using the fact that $R=a r$ in the second term, we can integrate
the above equation. We find
\begin{equation}\label{Frgb3}
H^2+\frac{k}{a^2}-\gamma
\left(H^2+\frac{k}{a^2}\right)^2=\frac{8\pi}{3}(\rho+\rho_{\Lambda}),
\end{equation}
where $\gamma=5 l^2_{0}/2$ and we have used relation
(\ref{radius}) in the last step.  This result coincides with the
one obtained from the first law of thermodynamics in Eq.
(\ref{Fried4}). This further supports the viability of the
Padmanabhan's perspective of emergence gravity.
\section{Generalized second law of thermodynamics} Now we are
going to check whether the obtained entropy associated with the
apparent horizon, given in Eq. (\ref{St}) can satisfy the
generalized second law of thermodynamics. For our Universe, which
is currently experiencing a phase of accelerated expansion, the
generalized second law of thermodynamics has been investigated in
\cite{Wang1,wang2,SheyGSL}.

Using Eq. (\ref{Fried1}) as well as the continuity equation, we
get
\begin{equation} \label{dotR}
\dot{R}= 4\pi R^3 H (\rho+p)
\left(1+\frac{l_0^2}{R^2}\right)^{3/2}.
\end{equation}
It is a matter of calculations to show that
\begin{eqnarray}\label{TSh1}
T_{h} \dot{S_{h}}&=&4\pi H R^3 (\rho+p)\left(1-\frac{\dot
{R}}{2HR}\right).
\end{eqnarray}
Since in an accelerated universe with fantom regime, the strong
energy condition may be violated, namely $\rho+p<0$. This implies
that the second law of thermodynamics may break down,
$\dot{S_{h}}<0$. Therefore we consider the generalized second law
of thermodynamics. From the Gibbs equation we have \cite{Pavon2}
\begin{equation}\label{Gib2}
T_m dS_{m}=d(\rho V)+pdV=V d\rho+(\rho+p)dV,
\end{equation}
where $T_{m}$ and $S_m$ stand for the temperature and entropy of
the matter fields in the bulk, respectively. Assuming the thermal
equilibrium hypothesis, the temperature of the matter field in the
bulk and the horizon temperature are the same, namely $T_m\approx
T_h$ \cite{Pavon2}. This assumption is necessary to prevent the
flow of energy between the bulk and the boundary. The Gibbs
equation (\ref{Gib2}) can be easily written as
\begin{equation}\label{TSm2}
T_{h} \dot{S}_{m} =4\pi {R^2}\dot {R}(\rho+p)-4\pi {R^3}H(\rho+p).
\end{equation}
Adding Eqs. (\ref{TSh1}) and (\ref{TSm2}), after using Eq.
(\ref{dotR}), we reach
\begin{equation}\label{GSL3}
T_{h}( \dot{S_{h}}+\dot{S_{m}})=8 \pi^2 H R^{5}(\rho+p)^2
\left(1+\frac{l_0^2}{R^2}\right)^{3/2}.
\end{equation}
{From this equation we have $\dot{S_{h}}+\dot{S_{m}}\geq 0$, which
implies that the generalized second law of thermodynamics hold
when the entropy associated with the apparent horizon of FRW
universe consists zero-point length correction terms as given in
Eq. (\ref{St}). However, for other entropy expressions this issue
should be checked separately.}
\section{Cosmological implications \label{cosmology}}
In this section, we would like to study cosmological consequences
of the modified Friedmann equations. For simplicity, we consider a
flat universe, although one can generalize the arguments to a
nonflat universe.
\subsection{Matter dominated era}
Let us begin with the case where our Universe is dominated with
pressureless matter and the cosmological constant is equal to
zero. From the continuity equation $\dot{\rho}_m+3H \rho_m=0$, we
have $\rho_m=\rho_{m,0} a^{-3}$. Dividing both sides of Eq.
(\ref{Fried6}) by $H_0^2$, we find
\begin{equation} \label{Frim1}
-\beta
\left(\frac{H}{H_0}\right)^4+\left(\frac{H}{H_0}\right)^2-\Omega_{m,0}
a^{-3}=0,
\end{equation}
where we have used definition $\Omega_{m,0}={8\pi \rho_{m,0}}/(3
H_0^2 )$ and $\beta\equiv\alpha H_0 ^2=3l_0^2 H_0 ^2/4\ll 1$. Note
that at the present time ($H_0=H)$ and ($a=a_0=1)$, Eq.
(\ref{Frim1}) can be written as
\begin{equation} \label{Frim11}
\Omega_{m,0}+\beta=1,
\end{equation}
Solving Eq. (\ref{Frim1}), we find
\begin{equation} \label{Frim2}
\left(\frac{H}{H_0}\right)^2=-\frac{1}{2\beta}\left[-1+
\sqrt{1-4\beta \Omega_{m,0} a^{-3}}\right].
\end{equation}
Since $\beta$ is very small, we can expand the RHS of the above
equation up to the linear term in $\beta$. The result is
\begin{equation} \label{Frim3}
\frac{H^2}{H_0^2}=\Omega_{m,0} a^{-3}\left(1+\beta \Omega_{m,0}
a^{-3}\right)+\mathcal{O}(\beta^2).
\end{equation}
If we define, as usual, the redshift parameter as $a^{-1}=1+z$, we
can rewrite Eq. (\ref{Frim3}) as
\begin{equation} \label{Frim4}
\frac{H^2(z)}{H_0^2}=\Omega_{m,0} (1+z)^{3}\Big{\{}1+\beta
\Omega_{m,0} (1+z)^{3}\Big{\}}+\mathcal{O}(\beta^2).
\end{equation}
Let us calculate Eq.(\ref{Frim4}) at the present time ($z=0$).
Using Eq. (\ref{Frim11}), we have
\begin{eqnarray} \label{Frim5}
\left(\frac{H}{H_0}\right)^2_{z=0}&=&\Omega_{m,0}+\beta
\Omega^2_{m,0}+\mathcal{O}(\beta^2)\nonumber\\
&=&(1-\beta)+\beta(1-\beta)^2\simeq1+\mathcal{O}(\beta^2),
\end{eqnarray}
where we have neglected the terms of $\mathcal{O}(\beta^2)$  and
higher order terms of $\beta$. We can also rewrite Eq. (\ref{Frim3}) as
\begin{equation} \label{Frim6}
\frac {da}{a \sqrt{\Omega_{m,0} a^{-3}\left(1+\beta \Omega_{m,0}
a^{-3}\right)}}=H_{0}dt.
\end{equation}
where we have used $H=(da/dt)/a$. By integrating the aforementioned
relation, we arrive at
\begin{equation}
\frac{2a^{3/2}}{3\sqrt{\Omega _{m,0}}}+\frac{\beta}{3} \sqrt{\frac{\Omega _{m,0}}{a^3}}=H_{0}t,  \label{atmd}
\end{equation}
Finally, we can obtain the explicit form of the scale factor as a
function of time. We find
\begin{equation} \label{am1}
a(t)=\Bigg{\{}\frac{9}{4}\Omega_{m,0}\left(H_0 ^2 t^2-\frac{4
\beta}{9}\right)\Bigg{\}}^{1/3},
\end{equation}
where the integration constant is assumed to be zero. In the
limiting case where $\beta=0$, one recover $a(t)\sim t^{2/3}$,
which is the well-known solution in standard cosmology for the
matter only Friedmann model.
\begin{figure}[H]
\includegraphics[scale=0.8]{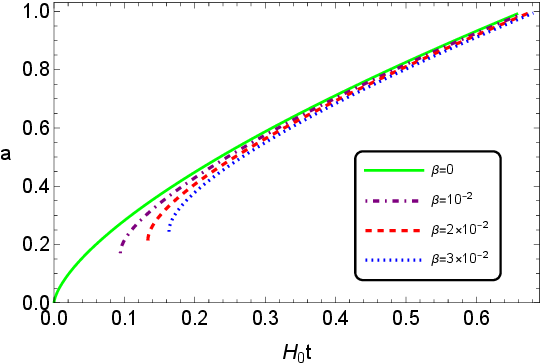}
\caption{The behavior of $a(t)$ in terms of $H_{0}t$ for matter-dominated era with different values of $%
\protect\beta $. Note that $\Omega _{m,0}=1-\protect\beta $.}
\label{fig1}
\end{figure}
In Fig. \ref{fig1}, we explore the behavior of the scale factor
$a(t)$ for different values of $\beta$. We see that in zero-point
length cosmology ($\beta\neq0 $), the scale factor lies below the
scale factor of standard cosmology, where $\beta=0$. In addition,
we observe that for a given value of $H_0 t$, the scale factor
decreases as $\beta $ increases. Note that for $\beta\neq0$, there
is a lower bound on $H_0 t$ in which $H_0 t > 2\sqrt{\beta}/3$
which leads to $t> l_0/\sqrt{3}$. Note that $c=1$ in our
discussion. Clearly, for $H_0 t< 2\sqrt{\beta}/3$ one needs to
consider the effects of quantum gravity and the classical general
relativity no longer valid.

Taking the time derivative of Eq. (\ref{Frim3}), after some
algebra, we obtain the deceleration parameter as
\begin{equation} \label{qm1}
q(z)=-1-\frac{\dot{H}}{H^2}=\frac{1}{2}+ \frac{3}{2} \times
\frac{\beta \Omega_{m,0} (1+z)^3 }{1+\beta \Omega_{m,0} (1+z)^3}.
\end{equation}
Eq. (\ref{qm1}) indicates that in zero-point length cosmology and
for a matter dominated universe we have always a decelerated
universe where $q>1/2$.
\begin{figure}[H]
\includegraphics[scale=0.8]{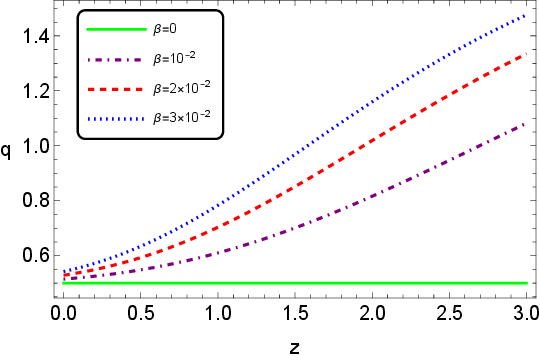}
\caption{The behavior of the deceleration parameter $q$ in terms
of redshift parameter $z$ for matter-dominated era with different
values of $\protect\beta $. Note that $\Omega
_{m,0}=1-\protect\beta $.} \label{fig2}
\end{figure}
In Fig. \ref{fig2}, we plot the deceleration parameter $q$ versus
redshift parameter $z$. We see that $q(z)$ is always greater than
$1/2$ during the history of the universe.

We can also estimate the age of the Universe for the matter only
dominated Universe, when the zero-point length is taken into
account. The age of the Universe is defined as
\begin{equation} \label{Age1}
t_0=\int_{0}^{\infty}{\frac{dz}{(1+z)H(z)}}.
\end{equation}
Inserting $H(z)$ from (\ref{Frim4}), we have
\begin{equation} \label{Age2}
t_0=\frac{1}{H_0}\int_{0}^{\infty}{\frac{dz(1+z)^{-1}
}{\sqrt{\Omega_{m,0} (1+z)^{3}\left[1+\beta \Omega_{m,0}
(1+z)^{3}\right]}}}.
\end{equation}
If we change the variable as $(1+z)^{-1}=x$, the above expression
can be written as
\begin{eqnarray} \label{Age3}
t_0&=&\frac{1}{H_0}\int_{0}^{1}{\frac{x
dx}{\sqrt{\Omega_{m,0}x\left[1+\beta \Omega_{m,0}
x^{-3}\right]}}}\nonumber\\
&=&\frac{2}{3 H_0 \sqrt{\Omega_{m,0}}}
\left(\sqrt{1+\beta\Omega_{m,0}}-\sqrt{\beta\Omega_{m,0} }
\right).
\end{eqnarray}
Using the fact that $\beta=1-\Omega_{m,0}$, we can rewrite the
final form of the age as
\begin{eqnarray} \label{Age4}
t_0=\frac{2}{3 H_0}
\left(\sqrt{1+\Omega_{m,0}^{-1}-\Omega_{m,0}}-\sqrt{1-\Omega_{m,0}}\right).
\end{eqnarray}
For $\beta=0$, we have $\Omega_{m,0}=1$ and one recover $t_0=2/(3
H_0)$, which is well-known result in standard cosmology for the
matter dominated universe.
\begin{table}[H]
\begin{center}
\begin{tabular}{|c{1.5cm}|c{1.4cm}|c{1.4cm}|c{1.4cm}|c{1.4cm}|}
\hline
$\beta$ & $0$ & $10^{-2}$ & $2\times10^{-2}$  & $3\times10^{-2}$   \\ \hline
$H_0 t_{0}$ & $0.66$ & $0.67$ & $0.68$  & $0.69$   \\ \hline
\end{tabular}%
\end{center}
\caption{The values of $H_0 t_0$  for a
matter-dominated universe and different values of $\protect\beta$.}
\label{table1m}
\end{table}

In Table \ref{table1m}, we present the age of the universe for
different values of $\beta$ within the framework of zero-point
length cosmology in a matter-dominated universe. Compared to
standard cosmology, we observe that the age of the universe
increases.

Substituting the scale factor from (\ref{am1}) into $\rho_{m}(t)=
\rho_{m,0}\,  a^{-3}$, we obtain
  \begin{align}
 \frac{\rho_{m}(t)}{\rho_{m,0}}=\frac{4}{9(1-\beta)}\, \frac{1}{\left(H_0 ^2 t^2-\frac{4
 \beta}{9}\right)},
 \end{align}
where we have used $\beta=1-\Omega_{m,0}$.

\begin{figure}[H]
\includegraphics[scale=0.9]{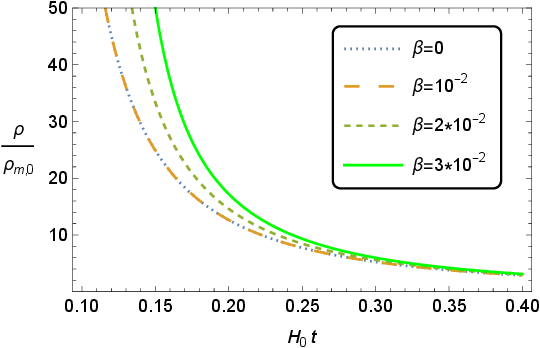}
\caption{The behavior of energy density $\rho(t)$ in terms of
$H_{0}t$ for matter-dominated era and different values of $\beta
$.} \label{fig3}
\end{figure}
Fig. \ref{fig3} illustrates the behavior of the energy density
$\rho(t)$ for different values of $\beta$. We observe that in the early epoch of the universe as $\beta$ increases, the energy
density increases too. Since $\rho _m>0$, we must have $H_0
t>2\sqrt{\beta}/3$. Thus, $H_0 t$ cannot go to zero due to the
presence of zero-point length.
\subsection{Radiation dominated era}
It is generally believed that, following inflation, our universe
underwent a radiation-dominated era during which most particles
behaved relativistically. In this case the continuity equation
implies $\rho_{r}=\rho_{r,0} a^{-4}$ and the Friedmann equation
(\ref{Fried6}) can be rewritten as
\begin{equation} \label{Frad1}
-\beta
\left(\frac{H}{H_0}\right)^4+\left(\frac{H}{H_0}\right)^2-\Omega_{r,0}
a^{-4}=0.
\end{equation}
Solving the above equation and keeping the terms up to
$\mathcal{O}(\beta)$, we find
\begin{equation} \label{Frad2}
\frac{H^2}{H_0^2}= \Omega_{r,0} a^{-4}\left(1+\beta \Omega_{r,0}
a^{-4}\right)+\mathcal{O}(\beta^2).
\end{equation}
In terms of the redshift parameter, we have
\begin{equation} \label{Fradi3}
\frac{H^2(z)}{H_0^2}= \Omega_{r,0} (1+z)^{4}\Big{\{}1+\beta
\Omega_{r,0} (1+z)^{4}\Big{\}}+\mathcal{O}(\beta^2).
\end{equation}
We can also rewrite Eq. (\ref{Frad2}) as
\begin{equation} \label{Frad3}
\frac {da}{a \sqrt{\Omega_{r,0} a^{-4}\left(1+\beta \Omega_{r,0}.
a^{-4}\right)}}=H_{0}dt.
\end{equation}
Integrating the aforementioned relation, we arrive at
\begin{equation}
\frac{a^2}{2\sqrt{\Omega _{r,0}}}+\frac{\beta}{4} \sqrt{\frac{\Omega _{r,0}}{a^{4}}}=H_{0}t,  \label{atmd}
\end{equation}
Solving for the scale factor, we find
\begin{equation} \label{ar2}
a(t)=\Big{\{}4 \Omega_{r,0} \left(H_0^2 t^2 -\frac{\beta}{4}\right
)\Big{\}}^{1/4},
\end{equation}
where the integration constant is set to zero. When $\beta=0$, one
recovers $a(t)\sim t^{1/2}$, which is an expected result in
standard cosmology for the radiation dominated epoch.
\begin{figure}[H]
\includegraphics[scale=0.8]{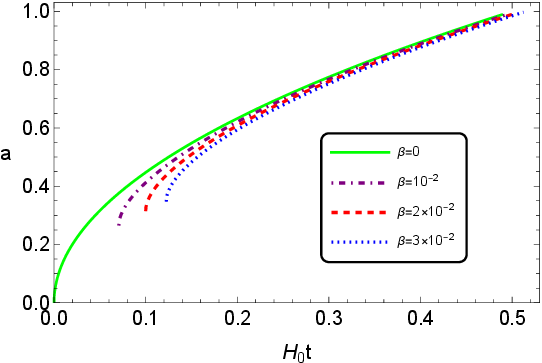}
\caption{The behavior of scale factor $a(t)$ vs $H_{0}t$ for radiation-dominated era with different values of $%
\protect\beta $. Note that $\Omega _{r,0}=1-\protect\beta $.}
\label{fig4}
\end{figure}
In Fig. \ref{fig4}, we plot the behavior of the scale factor
$a(t)$ for different values of $\beta$ in the radiation dominated
epoch. We observe that at each time as $\beta $ increases the
scale factor decreases. This indicates that the radius of the
universe is smaller compared to the standard cosmology. Besides,
for zero-point length cosmology the scale factor does not go to
zero for $H_0 t> \sqrt{\beta}/2$. This may show removing the
initial singularity of the Universe. To become more confident, we
calculate Ricci scalar and Kretschmann invariant which are given by
$R= g^{\mu \nu } R_{\mu \nu }$, and $K=R_{\mu \nu \alpha\rho}
R^{\mu \nu \alpha\rho } $, respectively. For a FRW metric they are
given by
\begin{align}  \label{kentropy2}
R&=\frac{6( a \ddot{a}+\dot{a}^2)}{a^2},\nonumber\\
K&=\frac{12(  \ddot{a}^2 a^2+\dot{a}^4)}{a^4},
\end{align}
where the "dot" denotes the derivative with respect to time. Substituting $a(t)$ from (\ref{ar2}) into the above expressions, we obtain
\begin{align}  \label{kentropy2}
R&=-\frac{{12  \beta  H_0^2}}{{(4H_0^2  t^2 - \beta)^2}},\nonumber\\
K&=\frac{{48\,(8 H_0^4  t^4 + 4 H_0^2  \beta  t^2 + \beta^2)
H_0^4}}{{(4 H_0^2  t^2 - \beta)^4}}.
\end{align}
For the domain $H_0 t> \sqrt{\beta}/2$, the Ricci scalar and the
Kretschmann invariant both have finite values and do not diverge.
On the other hand, in the limit $t\rightarrow 0$ we obtain the
finite values $ R=-{12 H_0^2}/{\beta}$ and $K={48 H_0^4}/{
\beta^2}$ for $\beta\neq0$. This implies the avoidance of an
initial singularity in the cosmological model due to the
zero-point length. Such a result is to be expected since the
zero-point length removes the singularity in the potential.

\begin{figure}[h]
\includegraphics[scale=0.8]{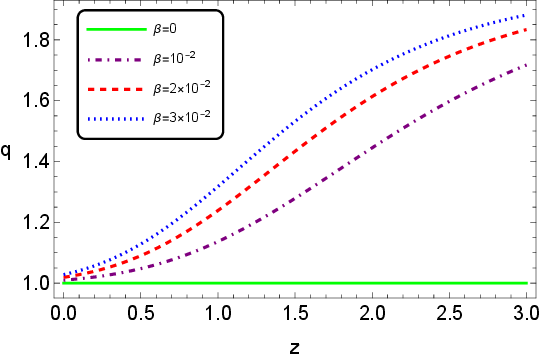}
\caption{The behavior of deceleration parameter $q$ vs redshift
$z$ for radiation-dominated era with different values of
$\protect\beta $. Note that $\Omega _{r,0}=1-\protect\beta $.}
\label{fig5}
\end{figure}

Furthermore, one can study the deceleration parameter which is given by
\begin{equation} \label{qr1}
q(z)=-1-\frac{\dot{H}}{H^2}=1+ \frac{\beta \Omega_{r,0} (1+z)^4
}{1+\beta \Omega_{r,0} (1+z)^4}.
\end{equation}
This quantity is shown to be always positive, indicating that our
Universe was undergoing a decelerated phase during the
radiation-dominated era (see Fig. \ref{fig5}).

Let us estimate the age of the Universe for a radiation-only
dominated era, taking into account the zero-point length. To do
so, we can substitute $H(z)$ from (\ref{Frim4}) into (\ref{Age1}),
yielding
\begin{equation} \label{Age2}
t_0=\frac{1}{H_0}\int_{0}^{\infty}{\frac{dz(1+z)^{-1}
}{\sqrt{\Omega_{r,0} (1+z)^{4}\left[1+\beta \Omega_{r,0}
(1+z)^{4}\right]}}}.
\end{equation}
Further by considering $(1+z)^{-1}=x$, the above expression
can be written as
\begin{eqnarray} \label{Age5}
t_0&=&\frac{1}{H_0}\int_{0}^{1}{\frac{x
dx}{\sqrt{\Omega_{r,0}\left[1+\beta \Omega_{r,0}
x^{-4}\right]}}}\nonumber\\
&=&\frac{1}{2 H_0 \sqrt{\Omega_{r,0}}}
\left(\sqrt{1+\beta\Omega_{r,0}}-\sqrt{\beta\Omega_{r,0} }
\right).
\end{eqnarray}
Using the fact that $\beta=1-\Omega_{r,0}$, we can write the expression for the age of a radiation-only dominated Universe as follows
\begin{eqnarray} \label{Age7}
t_0=\frac{1}{2 H_0}
\left(\sqrt{1+\Omega_{r,0}^{-1}-\Omega_{r,0}}-\sqrt{1-\Omega_{r,0}}\right).
\end{eqnarray}
In particular if we set $\beta=0$, we have $\Omega_{r,0}=1$, then one can recover $t_0=1/(2
H_0)$. This is an expected result in standard cosmology for the
radiation dominated Universe.
\begin{table}[h]
\begin{center}
\begin{tabular}{|c{1.4cm}|c{1.4cm}|c{1.4cm}|c{1.4cm}|c{1.4cm}|}
\hline
$\beta$ & $0$ & $10^{-2}$ & $2\times10^{-2}$  & $3\times10^{-2}$   \\ \hline
$H_0 t_{0}$ & $0.500$ & $0.505$ & $0.510$  & $0.515$   \\ \hline
\end{tabular}
\end{center}
\caption{The values of $H_0 t_0$ for different values of $\protect\beta$ in
radiation-dominated Universe}
\label{table2}
\end{table}
In Table \ref{table2}, we present the age of the Universe for different values of $\beta$. It can be observed that the age problem in standard cosmology can be mitigated.

By means of  Eq. (\ref{ar2}) and the relation $\rho_{r}(t)= \rho_{r,0}\,  a^{-4}$, we arrive at
\begin{align}
\frac{\rho_{r}(t)}{\rho_{r,0}}=\frac{1}{4}\, \frac{1}{ \Omega_{r,0} \left(H_0^2 t^2 -\frac{\beta}{4}\right)},
\end{align}
where $\Omega_{r,0}=1-\beta$. It follows that
\begin{align}
\frac{\rho_{r}(t)}{\rho_{r,0}}=\frac{1}{4}\, \frac{1}{(1-\beta) \left(H_0^2 t^2 -\frac{\beta}{4}\right)}
\end{align}
\begin{figure}[H]
\includegraphics[scale=0.9]{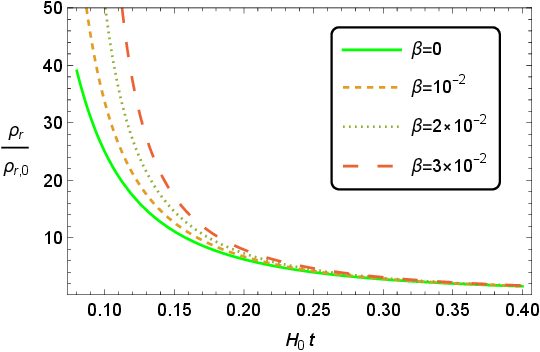}
\caption{The behavior of the energy density $\rho_{r}(t)$ vs
$H_{0}t$  for the radiation-dominated era with different values of
$\protect\beta $. Note that $\Omega _{r,0}=1-\protect\beta $.}
\label{fig6}
\end{figure}
Fig. \ref{fig6} illustrates the behavior of the energy density
$\rho_{r}(t)$  for different values of $\beta$ in a
radiation-dominated universe.  We observe that in the early
epoch, as $\beta$ increases, the energy density of the Universe
also increases. According to the equation $H_0 t>\sqrt{\beta}/2$,
we find that $\rho _r>0$, which is depicted in the Fig. \ref{fig6}.

Table \ref{comparison} compares our model's predictions for the
scale factor and age of the universe with standard and LQC models.
The $t^{1/4}$ scaling in the radiation era (Eq. 31) leaves
imprints on primordial gravitational waves, distinguishable via
future LISA observations \cite{LISA}.
\begin{table}[h]
    \begin{center}
        \begin{tabular}{|c{1.6cm}|c{1.6cm}|c{1.6cm}|c{1.6cm}|}
            \hline
            $Model$ & $a_r(t)$ & $a_m(t)$  & $Age (Gyr)$   \\ \hline
            $Standard$  & $t^{1/2}$ & $t^{2/3}$  & $13.8$   \\ \hline
            $LQC$  & $t^{1/3}$ & $t^{2/5}$  & $\sim14.2$   \\ \hline
            $Zeropoint$  & $t^{1/4}$ & $t^{1/3}$  & $14.0-14.5$   \\ \hline
        \end{tabular}
    \end{center}
    \caption{Comparison of cosmological predictions across models.}
    \label{comparison}
\end{table}
\subsection{Multiple component universe}
Let us now focus on the general case using the modified Friedmann equation, which can be written as
\begin{equation} \label{Frt11}
-\beta
\left(\frac{H}{H_0}\right)^4+\left(\frac{H}{H_0}\right)^2=\Omega_{m,0}
a^{-3}+\Omega_{r,0} a^{-4}+\Omega_{\Lambda,0}.
\end{equation}
If we neglect the contribution from radiation, which is small compared to matter and vacuum energy, we have the following relation
\begin{equation} \label{Frt2}
-\beta
\left(\frac{H}{H_0}\right)^4+\left(\frac{H}{H_0}\right)^2-(\Omega_{m,0}
a^{-3}+\Omega_{\Lambda,0})=0.
\end{equation}
Now, by solving the above equation for small $\beta$, we obtain
\begin{equation} \label{Frt3}
\frac{H^2}{H_0^2}=(\Omega_{m,0}
a^{-3}+\Omega_{\Lambda,0})\left[1+\beta (\Omega_{m,0}
a^{-3}+\Omega_{\Lambda,0})\right]+\mathcal{O}(\beta^2).
\end{equation}
Specifically, in terms of the redshift parameter, we have
\begin{eqnarray} \label{Frt4}
\frac{H^2(z)}{H_0^2}&=&
[\Omega_{m,0}(1+z)^{3}+\Omega_{\Lambda,0}]\nonumber \\
 && \times \left[1+\beta (\Omega_{m,0}
(1+z)^{3}+\Omega_{\Lambda,0})\right].
\end{eqnarray}
On the other hand, we can also rewrite Eq. (\ref{Frt3}) as
\begin{equation} \label{Frt5}
\frac {da}{a \sqrt{(\Omega_{m,0} a^{-3}+\Omega_{\Lambda,0})+\beta
(\Omega_{m,0} a^{-3}+\Omega_{\Lambda,0})^2}}=H_{0}dt.
\end{equation}

\begin{figure}[H]
\includegraphics[scale=0.8]{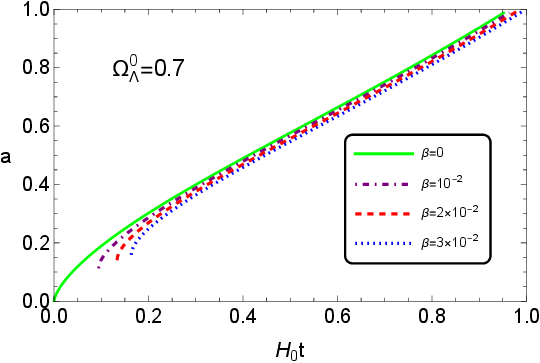}
\caption{We show the behavior of the scale factor $a(t)$ versus
$H_{0}t$ for a multiple-component Universe with different values
of $\protect\beta $. Note that $\Omega _{m,0}=1-\Omega
_{\Lambda,0}-\protect\beta $.} \label{fig7}
\end{figure}
Fig. \ref{fig7} illustrates the behavior of the scale factor
$a(t)$ for different values of $\beta$ in a multiple-component
universe. We observe that for a given value of $H_0 t$ as $\beta$
increases, the scale factor of the Universe decreases.
Consequently, the radius of the Universe is smaller compared to
that in standard cosmology. The plot shows that as time tends to
zero, the scale factor does not tend to zero.

Finally, the deceleration parameter is given by the expression
\begin{multline} \label{qmul}
q(z) =-1+\frac{3 \Omega_{m,0} (1+z)^3  }{2[\Omega_{m,0}
(1+z)^3+\Omega_{\Lambda,0}] }\\
 \times\frac{[1+2 \beta (\Omega_{m,0}
(1+z)^3+\Omega_{\Lambda,0})]}{[1+\beta (\Omega_{m,0}
(1+z)^3+\Omega_{\Lambda,0})]},
\end{multline}
\begin{figure}[H]
\includegraphics[scale=0.8]{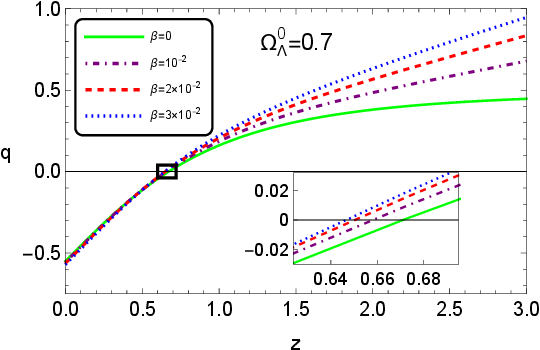}
\caption{The behavior of deceleration parameter $q$ vs redshift
$z$ for multiple-components Universe with different values of
$\protect\beta $. Note that $\Omega _{m,0}=1-\Omega _{\Lambda
,0}-\protect\beta $.} \label{fig8}
\end{figure}
Fig. \ref{fig8} shows the relationship between $q$ and $z$. Our
Universe experiences a phase transition from deceleration  ($q>0$)
to acceleration ($q<0$) during its history, and this transition
affects the formation and development of cosmic structures. We
observe that at lower redshifts, the transition from deceleration
to acceleration occurs. Moreover, it is evident that  in zero-point length cosmology, the onset of
cosmic acceleration happens later
compared to standard cosmology.

To calculate the age of the universe for a multi-component
universe by considering zero-point length, we substitute $H(z)$
from (\ref{Frt4}) in (\ref{Age1}), we have
\begin{equation}\label{Age8}
t_0=\frac{1}{H_0}\int_{0}^{\infty}{\frac{dz(1+z)^{-1}
}{\sqrt{X(1+\beta X)}}}.
\end{equation}
\begin{equation} \nonumber
X=\Omega_{m,0}
(1+z)^{3}+\Omega_{\Lambda,0}.
\end{equation}
By changing the variable as $(1+z)^{-1}=x$, the above expression
can be written as
\begin{multline} \label{Age5}
t_0=\frac{1}{H_0}\int_{0}^{1}{\frac{x
dx}{\sqrt{(\Omega_{m,0}x+\Omega_{\Lambda,0}x^4)\left[1+\beta( \Omega_{m,0}
x^{-3}+\Omega_{\Lambda,0})\right]}}}\nonumber\\
= \frac{1}{H_0} \left( \frac{2 \, \tanh^{-1} \left( \frac{\sqrt{\Omega_{\Lambda,0}}}{\sqrt{\Omega_{m,0} + \Omega_{\Lambda,0}}} \right)}{3 \sqrt{\Omega_{\Lambda,0}}} \right.
 + \frac{\beta  \sqrt{\Omega_{m,0} + \Omega_{\Lambda,0}}}{3}\\
 - \left.\frac{\beta}{3} \frac{\sqrt{\Omega_{\Lambda,0}} \, \sinh^{-1} \left( \frac{\sqrt{\Omega_{\Lambda,0}}}{\sqrt{\Omega_{m,0}}} \right)}{\sqrt{\Omega_{m,0}} \sqrt{1 + \frac{\Omega_{\Lambda,0}}{\Omega_{m,0}}}} \right)
\end{multline}
Note that in the general case  $\Omega _{m,0}=1-\Omega
_{\Lambda ,0}-\protect\beta $. For $\beta=\protect0$, we have  $\Omega _{m,0}=1-\Omega
_{\Lambda ,0} $. Thus, we can recover the age of multiple-component Universe in standard cosmology. Therefore, the above expression can be written as
\begin{equation}
t_0 = \frac{1}{H_0} \left(\frac{2 \, \tanh^{-1}  ( \sqrt{\Omega_{\Lambda,0}})} {3 \sqrt{\Omega_{\Lambda,0}}}\right).
 \end{equation}

\begin{table}[H]
\begin{center}
\begin{tabular}{|c{1.4cm}|c{1.4cm}|c{1.4cm}|c{1.4cm}|c{1.4cm}|}
\hline
$\beta$ & $0$ & $10^{-2}$ & $2\times10^{-2}$  & $3\times10^{-2}$   \\ \hline
$H_0 t_{0}$ & $0.96$ & $0.97$ & $0.98$  & $0.99$   \\ \hline
\end{tabular}%
\end{center}
\caption{The values of $H_0 t_0$ for a multi-component universe and various values of $\protect\beta$.}
\label{tablemul}
\end{table}
In Table \ref{tablemul}, we present the age of the Universe for
different values of $\beta$ for a multi-component universe, where
the contribution of radiation is negligible compared to other
components. It is observed that the age of the universe increases
in zero-point length cosmology compared to $\Lambda$CDM model.

Let us note that our model reduces tension between Planck and
SH0ES. From table \ref{tablemul} we see that for $\beta =
0.01$--$0.03$, we have $H_0 t_0 \approx 0.97$--$0.99$, aligning
with both CMB ($H_0 \sim 67$) and local ($H_0 \sim 73$) values if
$t_0$ is slightly larger. A modest $\beta \sim 0.02$ reconciles
Planck and SH0ES values within $1\sigma$, as the increased age
offsets the higher $H_0$.

We can also examine the total equation of state parameter which is
defined as
\begin{align}
\omega_{tot}=\frac{p_m+p_{\Lambda}}{\rho_m+\rho_{\Lambda} }.
\end{align}
Using the fact that $p_m=0$ and dividing both the numerator and
denominator by $\rho_{\Lambda}$ we arrive at
\begin{align}
\omega_{tot}=\omega_{\Lambda}\, \frac{\Omega_{\Lambda}}{\Omega_m+\Omega_{\Lambda} }.
\end{align}
Using $\Omega_m+\Omega_{\Lambda}=1-\beta $, we obtain
\begin{align}\label{key}
\omega_{tot}=\omega_{\Lambda}\, \frac{\Omega_{\Lambda}}{1-\beta}.
\end{align}
Using the definition $\Omega_{\Lambda}= \frac{8\pi
\rho_{\Lambda}}{3H^2} $ and $\rho_{\Lambda}=\Lambda/8 \pi$, we
find
\begin{align}\label{l}
\Omega_{\Lambda}= \frac{{\Lambda}}{3H^2}.
\end{align}
Substitute $H(z)$ from Eq. (\ref{Frt4}), we obtain
\begin{multline}\label{ome}
\Omega_{\Lambda}=\frac{{\Lambda}}{3H_{0}^2 }\\
\times\left[\frac{1}{(\Omega_{m,0}(1+z)^{3}+\Omega_{\Lambda,0})\left(1+\beta
(\Omega_{m,0} (1+z)^{3}+\Omega_{\Lambda,0})\right)}\right].
\end{multline}
 Combining relations (\ref{key}) and  (\ref{ome}) and using $\omega_{\Lambda}=-1$, we find
\begin{multline}
\omega_{tot}=\frac{-{\Lambda}}{3H_{0}^2 (1-\beta)}\\
\times\left[\frac{1}{(\Omega_{m,0}(1+z)^{3}+\Omega_{\Lambda,0})\left(1+\beta
(\Omega_{m,0} (1+z)^{3}+\Omega_{\Lambda,0})\right)}\right].
\end{multline}
\begin{figure}[H]
\includegraphics[scale=0.95]{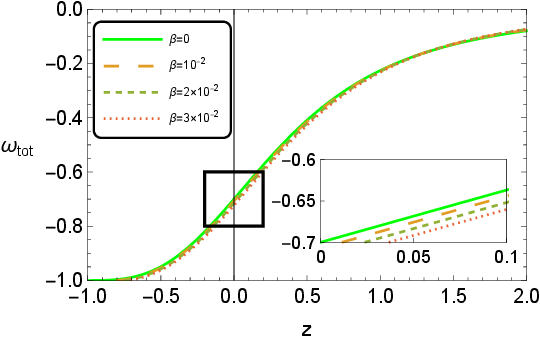}
\caption{The behavior of $\omega_{tot}$ vs redshift $z$ in a
multiple-components universe with different values of
$\protect\beta $.} \label{fig9}
\end{figure}
\begin{table}[H]
\begin{center}
\begin{tabular}{|c{1.4cm}|c{1.4cm}|c{1.4cm}|c{1.4cm}|c{1.4cm}|}
\hline
$\beta$ & $0$ & $10^{-2}$ & $2\times10^{-2}$  & $3\times10^{-2}$   \\ \hline
$\omega_{tot}$ & $-0.700$ & $-0.707$ & $-0.714$  & $-0.722$   \\ \hline
\end{tabular}
\end{center}
\caption{The values of $\omega_{tot}$ for present time ($z=0$) for different values of $\protect\beta$.}
\label{tabl4}
\end{table}
In Fig. \ref{fig9}, we plot the behavior of $\omega_{tot}$ for
different values of $\protect\beta$. We observe that at a given $z$,
the total equation of state parameter decreases with increasing
$\beta$. In Table \ref{tabl4}, we also present the total equation
of state parameter for different values of $\beta$ in a
multi-component universe at the present time ($z=0$). It is
observed that as $\beta$ increases, $\omega_{tot}$ decreases as
well.

A de Sitter universe is a solution to the Einstein field equations
of general relativity, named after Willem de Sitter. It models the
universe as spatially flat while neglecting ordinary matter,
resulting in dynamics dominated by the cosmological constant.
Therefore, by ignoring the contributions from radiation and matter
in Eq. (\ref{Frt11}), we obtain
\begin{equation} \label{Frad1}
-\beta
\left(\frac{H}{H_0}\right)^4+\left(\frac{H}{H_0}\right)^2-\Omega_{\Lambda,0}
=0.
\end{equation}
Solving the above equation and retaining terms up to the linear term in $\beta$, the result is
\begin{equation} \label{Fradi3}
\frac{H^2}{H_0^2}=\Omega_{\Lambda,0} +\beta \,{ \Omega_{\Lambda,0}}^2+\mathcal{O}(\beta^2).
\end{equation}
We can can rewrite the above equation as
\begin{equation} \label{Frad3}
\frac {da}{a \sqrt{\Omega_{\Lambda,0} +\beta \,{ \Omega_{\Lambda,0}}^2}}=H_{0}dt.
\end{equation}
Finally, we can obtain the explicit form of the scale factor as a function of time in a de Sitter universe. This results in
\begin{equation} \label{Frad3}
a(t)=\exp \left[  \sqrt{\Omega_{\Lambda,0}}\sqrt{1+\beta \,{
\Omega_{\Lambda,0}}} \, H_{0}t \right].
\end{equation}
Using Eq. (\ref{l}), the final form of the scale factor can be
rewritten as
\begin{equation} \label{Frad3}
a(t)=\exp \left[  \sqrt{\frac{\Lambda}{3}}\sqrt{1+\beta \,{
\Omega_{\Lambda,0}}} \,t \right].
\end{equation}
In the limiting case when $\beta=0$, we recover the expression
$a(t)\sim \exp\left( \sqrt{\frac{\Lambda}{3}}\,t\right)$, which is
the well-known solution in standard cosmology for the de Sitter
Universe.
\begin{figure}[H]
\includegraphics[scale=0.99]{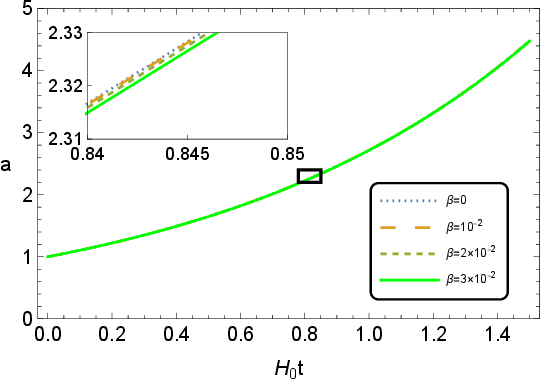}
\caption{The behavior of scale factor $a(t)$ vs $t$ for de Sitter Universe with different values of $\protect\beta    $. Note that $\Omega _{\Lambda,0}=1-\protect\beta $.}
\label{fig10}
\end{figure}
In Fig. \ref{fig10}, we study the behavior of the scale factor
$a(t)$ for different values of $\beta$ in a de Sitter universe. As
$\beta$ increases, the predicted age of the Universe also
increases compared to standard cosmology. Additionally, at each
point in time, the scale factor decreases as $\beta$ increases.
Consequently, the radius of the universe decreases compared to
standard cosmology.

Finally, for the deceleration parameter we have
\begin{equation} \label{qr1}
q(z)=-1-\frac{\dot{H}}{H^2}=-1.
\end{equation}
This value is always negative, indicating that the de Sitter
universe was underwent an accelerated phase.
\section{Closing remarks \label{Con}}
In order to address fundamental questions in modern cosmology,
exploring modified cosmology can enhance our understanding of the
universe. In this paper, by incorporating zero-point length
corrections to the gravitational potential and applying the
entropic force scenario, we obtained corrections to the entropy
associated with the horizon. In particular, we started from the
first law of thermodynamics, $dE=T_hdS_h+WdV$, on the apparent
horizon of the FRW universe, and obtained the modified Friedmann
equations. The implications of these corrections and their impact
on the generalized second law of thermodynamics are discussed.

We also derived the same modified Friedmann equations from a
different perspective, specifically through the concept of
emergent gravity. By calculating the number of degrees of freedom
in the bulk and on the boundary of the universe, we reached the
same result.

In the following sections of the paper, we analyzed in details the
cosmological implications of the modified Friedmann equations.
Specifically, we focused on single-component flat universes
dominated by matter, radiation, and de Sitter space, as well as
multi-component flat universes composed of pressureless matter and
a cosmological constant or dark energy. It is argued that in
modified cosmology inspired by zero-point length, the predicted
age of the universe is greater compared to standard cosmology.
Consequently, the age problem can be mitigated. Additionally, it
is observed that the scale factor decreases compared to standard
cosmology, resulting in a smaller radius of the universe. Our
findings also indicate that the transition from a decelerated
universe to an accelerated one occurred at lower redshifts
compared to standard cosmology, with the onset of cosmic
acceleration happening later.

Focusing on modified cosmology with only pressureless matter or a
radiation-dominated era, and in the absence of any dark energy,
results in a decelerating universe. This model cannot account for
an accelerated expansion without dark energy. However, it is shown
that as $ t \rightarrow 0$ both Ricci Scalar and Kretschmann
invariant have finite values and do not diverge for $\beta\neq0$.
Thus, the initial singularity of the universe can be mitigated by
the zero-point length.

As a final remark, we calculated the parameter of the equation of
state and the energy density of our universe and examined their
behavior for different values of  $\beta$. We observed that the
effects of $\beta$ are significant in the early universe but have
little impact on current times.

These findings not only demonstrate the influence of modified
entropy into the geometric component of field equations, but also
highlight the significant implications that arise from such
modifications. Such results challenge our current understanding of
the universe's evolution and provide valuable insights into its
dynamics.\\
\textbf{Acknowledgements:} { We are grateful to the respected
referees for constructive comments which helped us improve our
paper significantly. We also thank A. Dehyadeghari for the useful
discussion and valuable comments. The work of A. Sheykhi is based
upon research funded by Iran National Science Foundation (INSF)
under project No. 4022705.}

\end{document}